\documentclass{aa}  
\usepackage{xcolor}
\usepackage{graphicx}
\usepackage{txfonts}
\usepackage{lipsum}
\usepackage{subcaption}         
\usepackage{lscape}                                     
\usepackage{placeins}                                    
\usepackage{natbib}
\usepackage{hyperref}

\newcommand{\ud}{\mathrm{d}}

\begin{document}

   \title{Temperature-programmed desorption of SO$_2$ from water ice surfaces: Adsorption energy distributions}

    \author{F. Benoit\inst{1}
        \and A. B. Hacquard\inst{1}\inst{, 2} \and   {J.-H. Fillion\inst{1}} \and   {M. Bertin\inst{1}} \fnmsep
        }

   \institute{Sorbonne Université, CNRS, De la Molécule aux Nano-Objets: Réactivité, Interactions et Spectroscopies, MONARIS, 75005, Paris, France \and \emph{now at} IRAP, Université de Toulouse, CNRS, UT3, CNES, Toulouse, France\\
             \email{mathieu.bertin@sorbonne-universite.fr}}

   \date{Received September 30, 20XX}

  \abstract{Sulphur-bearing species play a key role in the chemical evolution of the interstellar medium and icy Solar System bodies such as the Jovian moons, yet the sulphur budget remains poorly constrained. Sulphur dioxide is considered one of the main sulphur reservoirs in icy environments, making its interaction with water ice surfaces highly relevant for astrochemical models.}
   {This work aims to extract adsorption energy distributions of SO$_2$ on water ice substrates as a relevant model for astrophysical environments to better constrain its thermal behaviour and solid-gas exchange for astrochemical simulations.} 
   {We performed a systematic experimental study of temperature-programmed desorption of SO$_2$ deposited on three types of surfaces -- polycrystalline gold, compact amorphous solid water (c-ASW), and crystalline water under ultra-high vacuum conditions -- to then extract, using a Polanyi-Wigner model, the adsorption energy distributions of SO$_2$ on each surface.}
   {We performed the extraction of the adsorption energy distribution of SO$_2$ deposited on water ice substrates. These exhibit a bimodal structure: a first physisorbed layer and a second, more strongly bound population. Only minor differences are observed between c-ASW and crystalline water ice in the behaviour of the distributions. We also provide mean values, most probable values, and width of the distribution. On average, the binding energy of SO$_2$ on water ice surface is 439 $\pm$ 41~meV.}
    {}


   \maketitle
   \nolinenumbers

\section{\label{sec:intro}Introduction}

Sulphur is a key element in the chemical evolution of the interstellar medium (ISM), where it participates in both gas-phase and solid-state reaction networks \citep{Mifsud_2021}. Although more than a dozen sulphur-bearing molecules have been identified in dense molecular clouds and protostellar environments, their total observed abundance remains far below the cosmic sulphur value inferred from diffuse clouds. This discrepancy, known as the missing sulphur problem, suggests that a substantial reservoir of sulphur may be locked in the icy mantles of interstellar grains \citep{Laas_2019}. Understanding the trapping, reactivity, and thermal evolution of sulphur species in ices is therefore essential for constraining the sulphur budget throughout the ISM. Moreover, sulphur is an important element on moons in the Solar System, most notably the Jovian moons, whose icy surfaces are constantly subjected to sulphur ion irradiation from Jupiter's magnetosphere \citep{cooper_energetic_2001}. Sulphur on the icy crusts of Ganymede, Callisto, and Europa is believed to be the source of the detection of sulphur-bearing molecules on their surfaces \citep{lane_evidence_1981,loeffler_thermallyinduced_2010, hodyss_ultraviolet_2019}. Both in the ISM and on the surface of icy moons, SO$_2$ is considered one of the main carriers of sulphur. In the ISM, its detection in the gas phase can be both associated with warmer regions, such as hot cores, where the sublimation of the ices take place, and be a tracer for shocked regions (e.g. \citet{taquet_seeds_2020, codella_solis_2021, van_gelder_joys_2024}). In both cases, it is thought to originate from the grains, a hypothesis further supported by JWST observations of SO$_2$ in the icy mantles  \citep{mcclure_ice_2023, rocha_jwst_2024}. 

In recent years, the adsorption and induced reactivity of SO$_2$ and as other sulphur-bearing molecules in the condensed phase have attracted considerable attention (e.g. \cite{loeffler_thermallyinduced_2010, schriver-mazzuoli_infrared_2003, kanuchova_thermal_2017, Bang_2017, Mifsud_2021, Mifsud_2023, martin-domenech_photodesorption_2025, basalgete_vacuum_2026}). These experimental studies have demonstrated that sulphur-bearing molecules and, in particular, SO$_2$ can undergo efficient hydrogenation, photodissociation, and thermally activated reactivity even at low temperatures in the condensed phase, providing precious data for modelling the formation and evolution of sulphur-bearing species in both the ISM and at the surface of icy moons. However, a precise reconstruction of the history and evolution of S-bearing molecules using astrochemical models critically depends on accurate numerical parameters, such as adsorption energies and desorption kinetics. These parameters drive adsorption, diffusion, and thermal exchanges between the solid and gaseous phase, thereby strongly impacting the locking and stability of sulphur in the solid phase, as well as its chemistry \citep{penteado_sensitivity_2017, Laas_2019}. In the case of sulphur bearing species, these adsorption energies remain poorly determined for astrochemically relevant substrates. Here, we aim at providing adsorption energies for SO$_2$ interacting with water ice surfaces as a simple model for SO$_2$ adsorption on interstellar dust grains or on the surface of Jovian moons. 

A single binding energy of 260$\pm$10~meV for SO$_2$ adsorption on a water ice has been reported by \cite{penteado_sensitivity_2017}. However, this value is an estimation extrapolated from non-quantified thermal desorption studies \citep{Collings_2004} via a very simple approach, i.e. only based on the relative desorption temperature compared to that of water. In addition, since the adsorption of molecules on disordered surfaces involves a variety of different binding sites and configurations, a single adsorption energy value cannot reliably model a given system, as adsorption energies usually span a wide range \citep{amiaud_2006, Zubkov_2007, bertin_nitrile_2017b, tinacci_theoretical_2022, tinacci_theoretical_2023, Minissale_2022}. By contrast, adsorption energy distributions provide a more accurate description of molecular adsorption and thermal behaviour. \cite{Nguyen_2024} have recently determined the adsorption energy of SO$_2$ on water ices using quantum chemistry approaches. Although only five adsorption sites were considered for each substrate, they obtained values ranging from $\sim$ 400 to 600~meV, i.e. significantly higher than the estimation of \cite{penteado_sensitivity_2017}. Here, we present a study specifically dedicated to experimentally determining the adsorption energy distributions of SO$_2$ adsorbed on water ice substrates.

To achieve this goal, we conducted a systematic experimental study of the temperature-programmed desorption (TPD) of SO$_2$ on compact amorphous solid water (c-ASW), crystalline H$_2$O ice, and -- for comparison -- polycrystalline gold surfaces. Although gold does not reproduce the properties of interstellar dust grains, it offers a clean and non-reactive reference substrate, making it particularly useful for isolating the intrinsic effect of water ice on SO$_2$ binding. By combining high-precision TPD measurements with a multi-bin reconstruction of the adsorption energy, we extracted the full distribution of binding energies experienced by SO$_2$ in both multilayer and submonolayer regimes. This comparative approach allows us not only to quantify the interaction strength of SO$_2$ with water ice, but also to assess the influence of morphological effects. Our results further reveal signatures consistent with molecular trapping and strong-site adsorption, highlighting the active chemical role of water ice in modulating SO$_2$ stability.

This paper is organised as follows. Section 2 presents the experimental setup and calibration procedures. Section 3 provides a qualitative analysis of the desorption profiles obtained for SO$_2$ on gold, amorphous, and crystalline water ices. Section 4 describes the methodology used to extract adsorption-energy distributions and discusses the multilayer and submonolayer regimes. Section 5 summarises the implications of our results for sulphur chemistry in interstellar ices.

\section{\label{sec:metho}Experimental method}

The experiments were performed using the SPICES (Surface Processes and ICES) setup {at Sorbonne Université (MONARIS)}, an ultra-high vacuum (UHV) chamber operating at a base pressure of $\sim 10^{-10}$ mbar. At its centre, a gold cubic sample holder is mounted on a closed-cycle helium cryostat, enabling cooling to 15~K and controlled heating of the substrate. The sample holder can be rotated and translated, enabling its gold surface to be oriented towards either the gas doser for ice growth or the quadrupole mass spectrometer (QMS) for desorption experiments. A thermocouple and a resistive heater are attached to the substrate to monitor and control its temperature with a precision of 0.1~K.

\begin{figure*}[h!]
   \centering
   \includegraphics[width=\hsize]{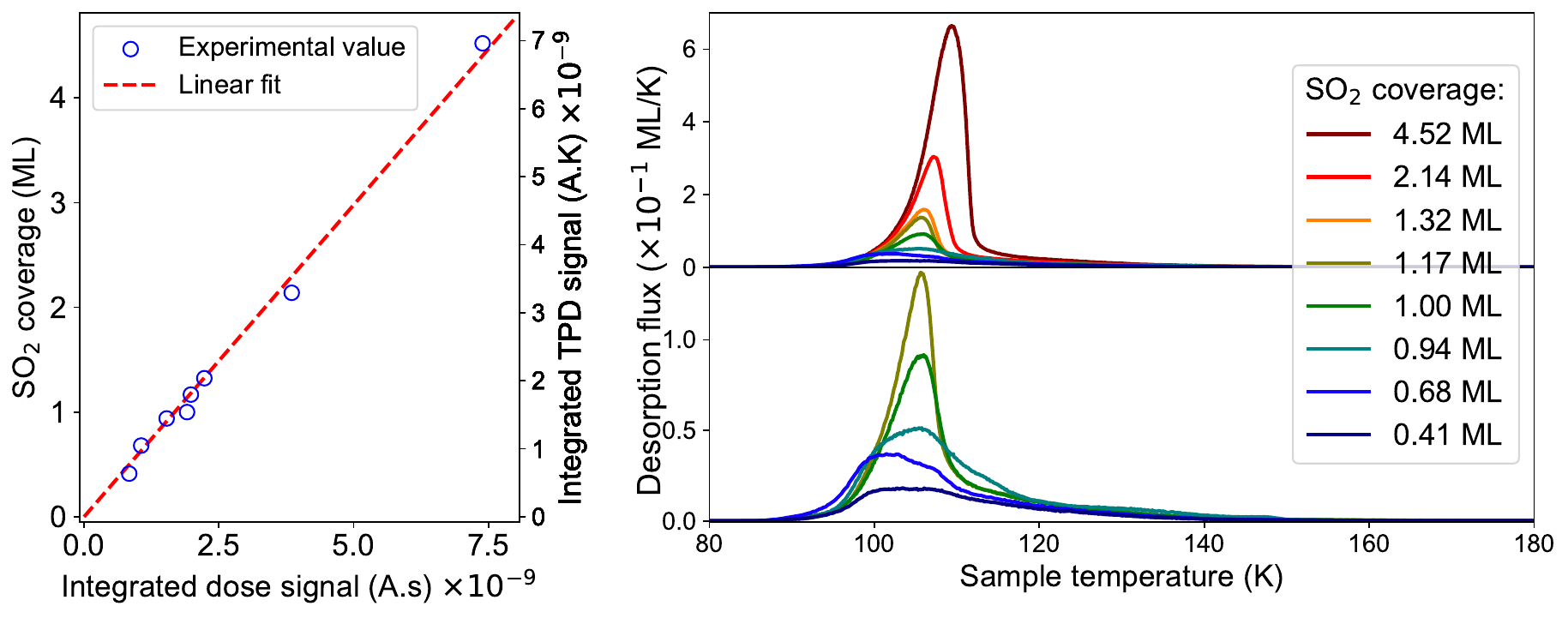}
   \caption{(Left) Integrated TPD signal as a function of the integrated dose signal for SO$_2$ deposited on gold at 80~K. The left axis indicates the ice thickness equivalence (expressed in monolayers) of the integrated TPD signal. The TPD experiments were performed with a linear heating rate of 12~K/min. The dashed red line represents the linear fit to the experimental data. The dose signal corresponds to the SO$_2$ mass signal measured at the doser outlet but not adsorbed on the substrate. (Right) Desorption flux as a function of sample temperature for different ice thicknesses. (Top right) All measured TPD curves. (Bottom right) Zoom-in on the TPD curves for coverages $\leq~1.17$~ML. The different leading edge observed for the TPD curve at 0.68~ML (blue curve) is associated with an experimental artefact in the temperature measurement, though it does not affect the coverage calibration.}
   \label{fig:calib_so2_au}
\end{figure*}

\begin{figure*}[h!]
   \centering
   \includegraphics[width=\hsize]{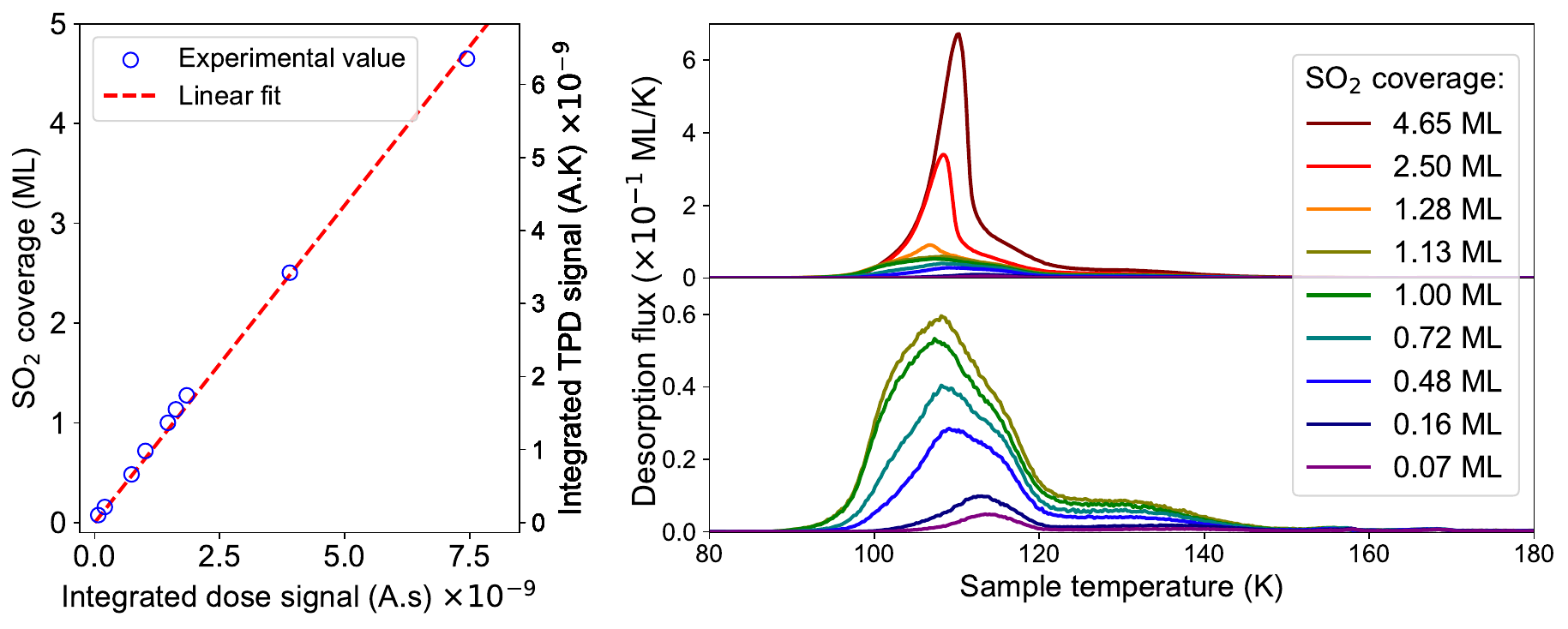}
   \caption{(Left) Integrated TPD signal as a function of the integrated dose signal for SO$_2$ deposited at 80~K on {20~ML of} c-ASW grown at 100~K on gold. The left axis indicates the ice thickness equivalence (expressed in monolayers) of the integrated TPD signal. The TPD experiments were performed with a linear heating rate of 12~K/min. The dashed red line represents the linear fit to the experimental data. The dose signal corresponds to the SO$_2$ mass signal measured at the doser outlet but not adsorbed on the substrate. (Right) Desorption flux as a function of the sample temperature for different SO$_2$ ice thicknesses on c-ASW. (Top right) All measured TPD curves. (Bottom right) Zoom-in on the TPD curves for coverages $\leq~1.13$~ML.}
   \label{fig:callib_so2_h2o}
\end{figure*}

Ices were grown by exposing the cold substrate to a partial pressure of gaseous molecules introduced into the chamber through a specially designed dosing tube positioned $1~$mm in front of the surface. This configuration enables the deposition of molecular ices with only a negligible, but still measurable, increase in background pressure. Good dose reproducibility relies on the precise mechanical positioning of the dosing tube relative to the sample. 

The number of molecules deposited on the cold substrate was estimated by using a QMS to monitor the fraction of molecules escaping from the dosing tube. After ice growth, TPD experiments were performed to verify that the integrated desorption flux was proportional to the integrated QMS signal recorded during dosing. This first calibration step, which establishes the proportionality between these two quantities, is shown in the left panels of Figs.~\ref{fig:calib_so2_au} and~\ref{fig:callib_so2_h2o}. The equivalence in monolayers is discussed later.

Here, three types of samples were {used}. First, (Fig.~\ref{fig:calib_so2_au}), SO$_2$ was deposited at $80~$K on a gold substrate. Then, (Fig.~\ref{fig:callib_so2_h2o}), SO$_2$ was deposited at $80~$K on c-ASW grown at $100~$K on a polycrystalline Au substrate. Lastly, SO$_2$ was deposited on crystalline water deposited at $140~$K on the gold substrate. In the third case, the crystalline character of the resulting water ice film was further verified during the TPD experiments, during which no additional crystallization was observed, as shown in Sect. 3 For both amorphous and crystalline water, 20~ML of H$_2$O (where 1~ML~$\approx$~$10^{15}$~molecules/cm$^{2}$) were deposited prior to SO$_2$ deposition. All TPD curves were obtained using a linear heating ramp of $\beta = 12~$K/min from 80~K to 200~K.  

Although deposition at 80~K differs from interstellar conditions, diffusion is expected to activate upon heating. Thus, the adsorption energy distributions we derive are representative of the system at the desorption temperatures, as is typically the case in TPD experiments \citep{Minissale_2022}. We therefore do not expect the initial deposition temperature to have an effect.

In the right panels of Figs.~\ref{fig:calib_so2_au} and~\ref{fig:callib_so2_h2o}, two desorption regimes can be distinguished. This behaviour can be understood by modelling the desorption flux as a kinetic process using the Polanyi–Wigner equation \citep{Redhead_1962}: 

\begin{equation}
\label{eq:pol_wig}
\Phi = - \frac{\mathrm{d} \theta}{\mathrm{d} T} = \frac{\nu}{\beta}\,\theta^N(T)\,\exp\left(-\frac{E_\mathrm{ads}}{kT}\right),
\end{equation}

where $\Phi$ is the desorption flux, $\theta$ the surface coverage as a function of temperature, $\nu$ the pre-exponential factor, $\beta$ the heating rate, $N$ the kinetic order, and $E_{\mathrm{ads}}$ the adsorption energy of SO$_2$ on the substrate. Here, the coverage is defined as the ratio between the number of deposited molecules and the number of available adsorption sites. According to this definition, two regimes can be distinguished: (i) a multilayer regime ($\theta > 1$) and (ii) a submonolayer regime ($\theta \leq 1$). In case (i), the kinetic order is $N = 0$ because the number of adsorbed molecules is effectively saturated; hence, desorption originates only from molecules located at the ice–vacuum interface, which remains constant. In case (ii), the number of molecules desorbing at a given time depends on the instantaneous coverage, resulting in a non-zero kinetic order ($N > 0$).

A series of TPD experiments were carried out at different coverages, as shown in the top right panels of Figs.~\ref{fig:calib_so2_au} and~\ref{fig:callib_so2_h2o}. The transition between different desorption regimes can be identified by a distinct break in slope in the rising part of the TPD curves, which corresponds to a change in desorption behaviour. This feature is clearly highlighted in the bottom right panels of the same figures. In Fig.~\ref{fig:calib_so2_au}, we assume that the monolayer desorption curve lies between the green (1~ML) and cyan (0.94~ML) curves. To ensure a consistent definition of the monolayer, we define the monolayer desorption as corresponding to the curve immediately preceding the break in slope when moving from lower to higher coverages. Based on this assumption, the corresponding TPD curve and the associated dosing signal are integrated, as shown in the left panel. For all investigated coverages (blue circles), a linear relationship is observed between the integrated TPD signal and the integrated dosing signal (linear fit in red). This linearity allows us to calibrate the SO$_2$ deposition for a given substrate -- namely, gold -- in Fig.~\ref{fig:calib_so2_au}. Since the calibration depends on both the adsorbed molecular species and the nature of the substrate, the same procedure was applied to a water substrate, as shown in Fig.~\ref{fig:callib_so2_h2o}. Using this method, we achieve a relative uncertainty of approximately 10\% on the derived coverages. 

The method used for adsorption energy estimation from experimental data relies on the reproducibility of the ice growth protocol. Therefore, both calibration and reproducibility are crucial. The purity of the species was also verified. For SO$_2$, we used a residual gas analyser to confirm the absence of isotopologues of $^{32}$S$^{16}$O$_2$ (Air Liquide, above $99.9\%$ purity). For H$_2$O, three freeze–pump–thaw cycles were performed to ensure purity.

\section{\label{sec:results}Desorption flux profiles}

\begin{figure}[h!]
\centering
\includegraphics[width=\hsize]{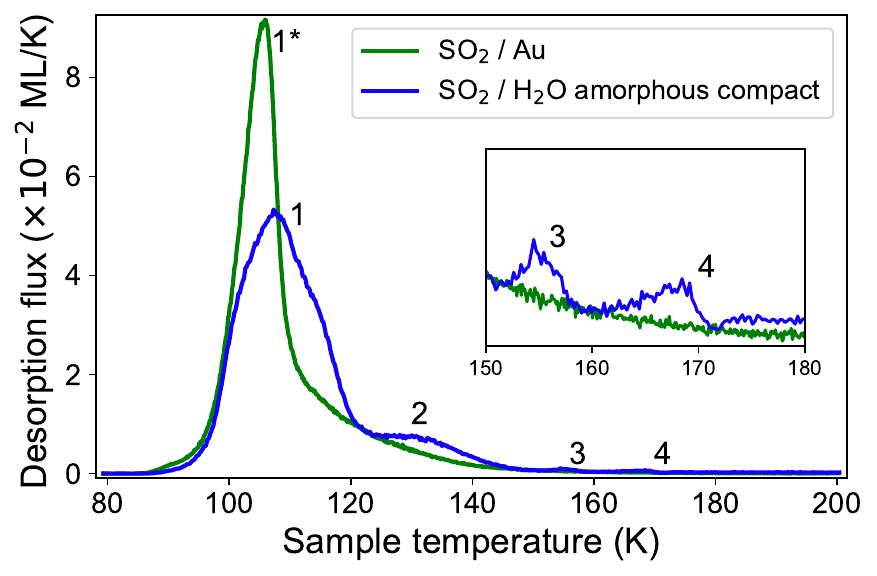}
\caption{TPD curves for 1~ML SO$_2$ deposited at 80~K on gold (green) and c-ASW (blue). The numbers indicate the different desorption peaks: a single peak for SO$_2$ on gold and four distinct peaks for SO$_2$ on c-ASW. The inset shows the TPD curves between 150 and 180~K, highlighting two additional, smaller desorption peaks that remain above the background level.}
\label{fig:comp_SO2_cASW}
\end{figure}

Figure \ref{fig:comp_SO2_cASW} shows TPD curves obtained for 1~ML of SO$_2$ deposited on bare gold and on 20~ML of c-ASW. Thermal desorption from Au (green curve) exhibits a single asymmetric peak at $\sim$~105~K (labelled 1*), with high temperature tails extending up to 140~K. A single component indicates that the adsorption of SO$_2$ on polycrystalline gold occurs via simple physisorption on the manifold of adsorption sites at the gold surface. When the same coverage is deposited on c-ASW (blue curve), the desorption features of SO$_2$ become more structured. The main desorption feature, labelled 1, broadens, shifts slightly towards higher energies, and appears multicomponent. In addition, a second desorption feature, centred at around 130~K, is also observed. Finally, two smaller contributions, 3 and 4, also appear at 155~K and 170~K.

By comparing the desorption curves from gold and c-ASW, one can conclude that peaks 1 and 1* are both associated with the desorption of weakly bound, physisorbed SO$_2$ interacting with the substrate. Adsorption energies of SO$_2$ on water and gold are expected to be different. This, along with varying orientations of surface water molecules creating diverse adsorption sites, can explain the broadening and shifting of peak 1. Peak 2 (120 - 140~K) is observed only when SO$_2$ is adsorbed on the water substrate and thus likely corresponds to more strongly bound molecules linked to specific interactions between H$_2$O and SO$_2$. In their study, \citet{Bang_2017} also identify these two desorption features. They conclude that peak 1 corresponds to the desorption of physisorbed surface SO$_2$ molecules, matching our observations. They also show that this desorption competes with the thermally activated reaction between SO$_2$ and water, which leads to the formation of HSO$_3^-$. They therefore propose that peak 2 is associated with SO$_2$ desorption from more strongly bound surface anions. It should be noted that thermal reactivity at 100-120~K in mixed SO$_2$-H$_2$O solids has also been highlighted by other studies using, for instance, infrared or UV spectroscopy \citep{loeffler_thermallyinduced_2010, hodyss_ultraviolet_2019, kanuchova_thermal_2017}.

Finally, peaks 3 and 4 correspond to the onset of c-ASW crystallization and water ice sublimation, respectively. These two contributions were already observed by \citet{Collings_2004}. Peak 3 can be attributed to thermally diffused SO$_2$ trapped in amorphous water during deposition or warm-up and released into the gas phase when water ice crystallizes. This well-known effect is referred to as volcano desorption \citep{Speedy_1996}. Peak 4 is also attributed to SO$_2$ molecules trapped within the crystalline ice bulk, released with water during co-desorption.

\begin{figure}[h!]
\centering
\includegraphics[width=\hsize]{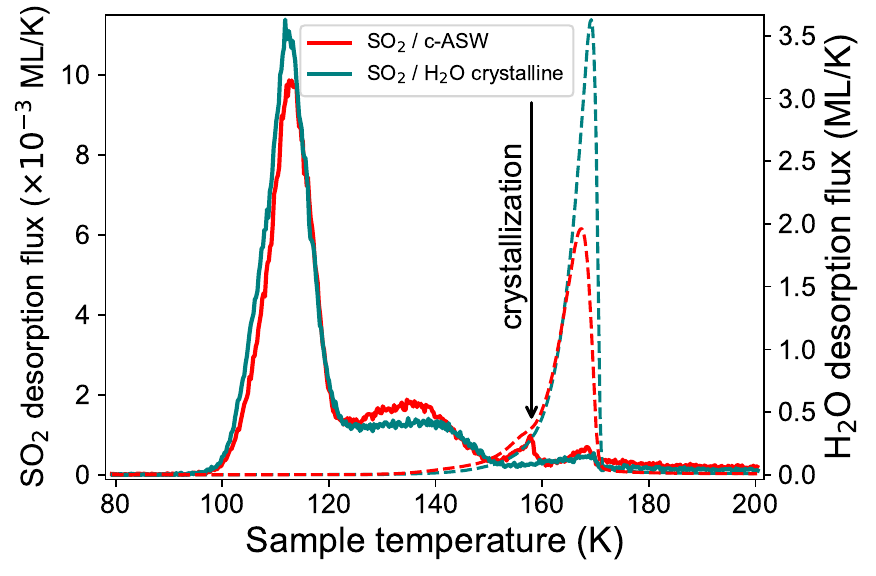}
\caption{TPD curves for 0.16~ML SO$_2$ deposited at 80~K on c-ASW (red) and crystalline H$_2$O (blue). The dashed lines represent the H$_2$O desorption flux from the same TPD experiments. The black arrow at 155~K indicates the c-ASW crystallization peak.}
\label{fig:comp_SO2_H2O_ac_cryst}
\end{figure}

These attributions are further confirmed by Fig. \ref{fig:comp_SO2_H2O_ac_cryst}, which shows TPD curves for 0.16~ML of SO$_2$ deposited at 80~K on 20~ML of either c-ASW (in red) or crystalline water (in cyan) substrates. Temperature-programmed desorption curves of the corresponding water ice are also shown as dashed lines. Peak 3 is indeed observed when SO$_2$ is deposited on amorphous water, but is absent when deposited on a crystalline substrate. Its observed desorption temperature matches amorphous ice crystallization, associated with the slope change at 155~K in the H$_2$O TPD curve. Peak 4 is observed for both water ices and matches the desorption curve of crystalline water, confirming co-desorption. Figure \ref{fig:comp_SO2_H2O_ac_cryst} also reveals no distinct differences in the two first SO$_2$ desorption contributions, indicating that the long-range molecular organization of the water surface does not strongly affect the physisorption or the thermal reactivity of the first SO$_2$ layer.

\section{\label{sec:dist_energies}  Extraction of quantitative adsorption energies}

The analysis of desorption flux profiles presented above highlights the diversity of SO$_2$–substrate interactions and their dependence on surface composition and morphology. It is essential to quantify these interactions by determining the adsorption energy distributions associated with the different desorption regimes. 
Such an analysis provides direct insight into the strength and nature of the binding between SO$_2$ molecules and their substrate and allows us to distinguish between multilayer and submonolayer adsorption regimes.

In the following, we discuss the adsorption energies derived for SO$_2$ from TPD data obtained on gold and water ices. 
Our discussion is divided into two main parts: 
(1) the multilayer regime, where the desorption kinetics reflect interactions between SO$_2$ molecules themselves, and (2) the submonolayer regime, where desorption is governed by SO$_2$–substrate interactions. 
For each case, we describe the fitting procedure, present the resulting energy distributions, and discuss the physical implications of the derived parameters.
\subsection{\label{sec:multi_regime}Multilayer regime}

As described in Sect.~\ref{sec:metho}, the number of molecules at the ice-vacuum interface remains constant during desorption in the multilayer regime.  
Consequently, the desorption flux is independent of total coverage, and the desorption order is $N = 0$.  
Under this condition, the Polanyi–Wigner equation can be expressed as
\begin{equation}
    \Phi = \frac{\nu}{\beta} \exp\left(-\frac{E_{\mathrm{ads}}}{kT}\right),
\end{equation}
which, after taking the logarithm, gives the linear relationship
\begin{equation}
    \ln \Phi = -\frac{E_{\mathrm{ads}}}{kT} + \ln\left( \frac{\nu}{\beta} \right),
    \label{eq:polanyi_multi}
\end{equation}
where $\Phi$ is the desorption flux, $\nu$ the pre-exponential factor, $\beta$ the heating rate, and $E_{\mathrm{ads}}$ the adsorption energy.

This linearized form enables a straightforward determination of both $E_{\mathrm{ads}}$ and the pre-exponential factor $\nu$ by fitting only the rising part of the TPD curve in the multilayer regime, where the assumption of zero-order desorption is valid.  
The slope of the linear fit yields $E_{\mathrm{ads}}$, while the intercept gives $\ln(\nu / \beta)$, from which $\nu$ can be directly calculated.  
The fits were obtained by minimising the squared deviation between the experimental data and the model described by Eq.~\ref{eq:polanyi_multi}.

\begin{figure}[h!]
   \centering
   \includegraphics[width=\hsize]{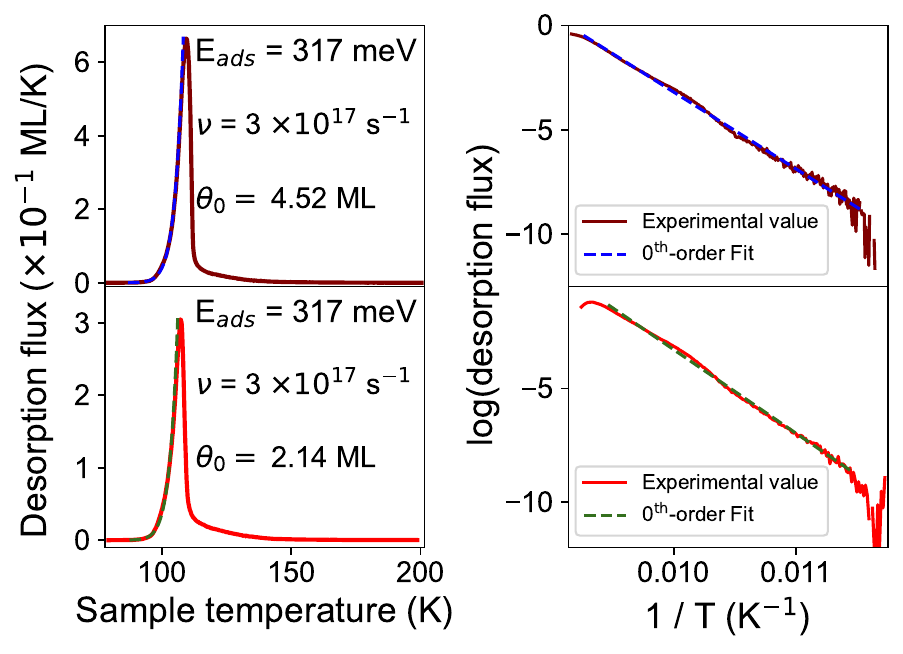}
   \caption{(Left) TPD curves of SO$_2$ deposited at 80~K on gold at different coverages (brown curve, 4.52~ML; red curve, 2.14~ML) in the multilayer regime, together with the corresponding zeroth-order fits (dashed blue and green lines, respectively).  
   (Right) Logarithm of the corresponding TPD curves as a function of the inverse sample temperature, with corresponding linear fits (dashed lines).  
   The adsorption energy and pre-exponential factor for SO$_2$ multilayer desorption are extracted from the slope and intercept of the linear regression.}
   \label{fig:multi_fit_so2}
\end{figure}

\begin{figure}[h!]
   \centering
   \includegraphics[width=\hsize]{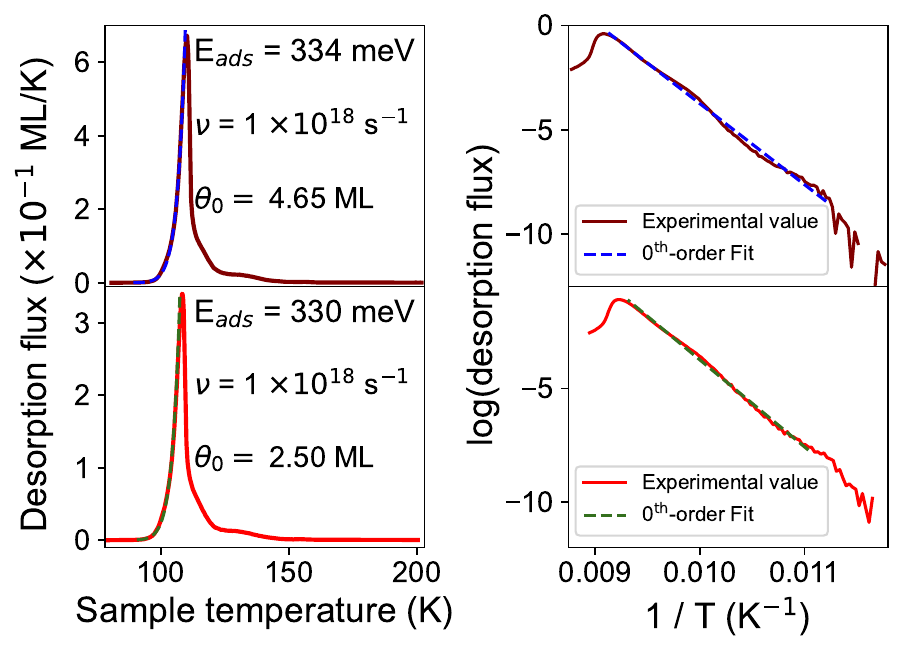}
   \caption{(Left) TPD curves of SO$_2$ deposited at 80~K on c-ASW at different coverages (brown curve, 4.65~ML; red curve, 2.50~ML) in the multilayer regime, together with the corresponding zeroth-order fits (dashed blue and green lines, respectively).  
   (Right) Logarithm of the corresponding TPD curves as a function of the inverse sample temperature, with corresponding linear fits (dashed lines).  
   The extracted values of $E_{\mathrm{ads}}$ and $\nu$ are reported in the panels.}
   \label{fig:multi_fit_so2_h2o}
\end{figure}

Figures~\ref{fig:multi_fit_so2} and~\ref{fig:multi_fit_so2_h2o} illustrate the fitting procedure for SO$_2$ desorption from gold and c-ASW, respectively.  
In the left panels, the experimental TPD curves (solid lines) are shown together with the corresponding zeroth-order fits (dashed lines) obtained using the optimized parameters.  
The right panels display the logarithm of the desorption flux as a function of the inverse temperature, along with the associated linear fits.  
Only the rising portion of each TPD curve was fitted, as the descending part corresponds to the transition to the submonolayer regime, which is not described by the zeroth-order model.  

In this regime, the fitting procedure can be delicate because low-temperature contributions to the desorption flux may slightly bias the estimation of the parameters ($E_\mathrm{ads}$ and $\nu$).  To minimise this effect, the fitting range was restricted to temperatures above $T = 90$~K, at which the desorption signal exceeds the background level by at least 10\% for all curves.  However, due to the fast heating ramp ($\beta = 12$~K/min), the signal in the 90–115~K temperature range not only originates from multilayer desorption but also includes a small contribution from monolayer desorption. Since all four multilayer curves correspond to coverages close to the monolayer regime ($\theta < 5$~ML), the monolayer and multilayer signals are of comparable magnitude, leading to a slight overlap. This overlap can slightly affect the extracted adsorption energy without changing the overall trend.  

From the fits, we obtain $E_\mathrm{ads} = 317 \pm 15$~meV and $\nu = 3\times10^{17 \pm 0.6 }$~s$^{-1}$ for SO$_2$ deposited on gold (Fig. \ref{fig:multi_fit_so2}), and $E_\mathrm{ads} = 332 \pm 15$~meV and $\nu = 1\times10^{18  \pm 0.6}$~s$^{-1}$ for SO$_2$ on c-ASW (Fig. \ref{fig:multi_fit_so2_h2o}). These values are in very good agreement with previous estimations of the desorption energy for thick SO$_2$ molecular solids found at 300$\pm30$~meV \citep{sandford_condensation_1993, schriver-mazzuoli_infrared_2003}. The small difference in the pre-exponential factors is consistent with the small temperature shifts observed between the TPD curves on the two substrates, as previously discussed. Since $\nu$ is directly related to the intercept of the linear fit, even a minor horizontal offset in temperature can lead to variations in its value.  
Overall, the results remain consistent within experimental uncertainties, confirming that multilayer SO$_2$ desorption is primarily governed by SO$_2$–SO$_2$ interactions rather than by the substrate. These multilayer adsorption energies provide a useful reference for comparison with the distributions derived in the submonolayer regime.  
Indeed, as observed in the TPD curves, the desorption peaks of SO$_2$ systematically occur at higher temperatures on both substrates when the coverage decreases, indicating that the mean or most probable adsorption energy in the submonolayer regime is higher than in the multilayer regime.  
This reflects the transition from SO$_2$–SO$_2$ dominated interactions to stronger SO$_2$–substrate interactions, which are analysed in the next section.

\subsection{\label{sec:submono_regime}Submonolayer regime}

In the submonolayer regime, the extraction of adsorption energies becomes more complex.  
Since the desorption order is no longer $N = 0$, three parameters must in principle be determined simultaneously: $N$, $\nu$, and $E_\mathrm{ads}$.  
To reduce the number of free parameters, we assumed a first-order desorption process ($N = 1$) and the presence of multiple adsorption sites, as discussed in \citep{Doronin_2015}.  
We further assumed that the prefactor $\nu$ does not vary significantly between sites.  
Under these assumptions, the Polanyi–Wigner equation can be rewritten as

\begin{equation}
    \Phi = -\frac{\ud \theta}{\ud T} = \frac{\nu}{\beta} \sum_i \theta_i(T) \, \exp\!\left(-\frac{E_i}{kT}\right),
    \label{eq:pol_submono}
\end{equation}

where the sum runs over the different adsorption sites $i$, each characterized by a coverage $\theta_i(T)$ and an adsorption energy $E_i$.  
In this formulation, only two parameters must be determined: $\nu$ and $E_\mathrm{ads}$.

  Following the transition state theory (TST) approach proposed by \citet{Tait_2005} and \citet{Minissale_2022}, the pre-exponential factor $\nu$ was estimated by assuming that the available thermal energy is insufficient for electronic and vibrational contributions to the partition functions to deviate from unity and that translational and rotational motions are frustrated in the adsorbed state. Under these assumptions, $\nu$ can be written as

\begin{equation}
    \nu = \frac{kT}{h} \, q_\mathrm{gas}^\mathrm{trans,2D} \, q_\mathrm{gas}^\mathrm{rot},
    \label{eq:prefactor_TST_final}
\end{equation}

with
\[
q_\mathrm{gas}^\mathrm{trans,2D} = A \left( \frac{2\pi m k T}{h^2} \right)^{3/2}, 
\qquad 
q_\mathrm{gas}^\mathrm{rot} = \frac{\sqrt{\pi}}{\sigma h^3} (8\pi^2 kT)^{3/2} \sqrt{I_x I_y I_z},
\]
where $k$ and $h$ are the Boltzmann and Planck constants, $T$ is the desorption peak temperature (the desorption window of SO$_2$ is narrow, with $\Delta T < 50~K$), $m$ is the molecular mass of SO$_2$, $A$ is the surface area per adsorbed molecule (assumed to be $10^{-19}~$m$^2$, corresponding to one monolayer), $\sigma$ is the rotational symmetry number, and $I_i$ ($i=x, y, z$) is the principal moment of inertia.

The numerical values used to compute $\nu$ are summarized in Table~\ref{table:prefactor}.

\begin{table}[h!]
\caption{Parameters used to calculate the prefactor $\nu$.}
\label{table:prefactor}
\centering
\begin{tabular}{c c c c c c c}
\hline \hline
\rule{0pt}{1.1em}
$T$ & $m$ & $I_x$ & $I_y$ & $I_z$ & $\sigma$ & $\nu$  \\ 
\hline
\rule{0pt}{1.1em}
107 & 64 & 16.9 & 48.6 & 32.8 & 2 & $6.7 \times 10^{17}$ \\
\hline
\end{tabular}
\tablefoot{$T$ is the peak desorption temperature (K), $m$ is the molecular mass (amu), and $I_x$, $I_y$, $I_z$ are the principal moments of inertia (amu·\AA$^2$).  {The values of $I_i$ are taken form \citep{nist_1997}.}}
\end{table}

With this prefactor, the experimental TPD curves presented in Figs.~\ref{fig:dist_fit_so2_au}, \ref{fig:dist_fit_so2_au_h2o}, and \ref{fig:dist_fit_so2_h2o_cryst} were fitted using a discrete multi-bin implementation of the Polanyi–Wigner law (Eq.~\ref{eq:pol_submono}).  
The desorption flux is expressed as the sum of independent desorption channels, each corresponding to a specific adsorption energy $E_i$ and fractional coverage $\theta_i$. For each energy bin, the desorption rate {is calculated by solving Eq. \ref{eq:pol_wig} assuming a first-order kinetics.} The total desorption flux is then $\Phi_\mathrm{tot}(T) = \sum_i \Phi_i(T)$. The fitting procedure minimises the quadratic deviation between the simulated and experimental desorption fluxes, under the constraint that the integrated areas (i.e. total desorbed amounts) are equal:
\[
\int \Phi_\mathrm{sim}(T) \, dT = \int \Phi_\mathrm{exp}(T) \, dT.
\]
The optimization variables are the fractional coverages $\theta_i$ associated with each energy bin, while $\nu$, $\beta$, and the energy grid $E_i$ are fixed parameters.  
Minimization was performed using a constrained least-squares algorithm, ensuring $\theta_i \ge 0$ and $\sum_i \theta_i(T_0) = \theta_0$.  
The resulting distribution $\theta(E)$ provides a direct numerical reconstruction of the adsorption energy landscape experienced by the SO$_2$ molecules in the submonolayer regime.

From the fitted adsorption energy distributions, the mean and standard deviation were computed as weighted moments of $\theta(E)$ according to Eq.~\ref{eq:moments}:
\begin{equation}
\langle E \rangle = \frac{\sum_i E_i , \theta_i}{\sum_i \theta_i}, \qquad
\sigma_E = \sqrt{ \frac{\sum_i (E_i - \langle E \rangle)^2 \theta_i}{\sum_i \theta_i} }.
\label{eq:moments}
\end{equation}

The energy grid was linearly spaced from 350~meV to 600~meV, with a step size of 5~meV.
Other intervals were tested, but the distributions either systematically went to zero outside this range or did not converge.
To avoid contributions from potential volcano-type sulphur desorption, which do not directly provide the SO$_2$ adsorption energy, the fitting procedure was restricted to temperatures below 150~K. 

It is important to note that this model introduces systematic uncertainties. The use of first order desorption kinetics reflects both the distribution of binding sites and the inhomogeneity of the local molecular environment, which would otherwise lead to an effective fractional order kinetic law that is difficult to interpret physically. The width of the resulting adsorption energy distribution therefore simultaneously accounts for both the effect of the different adsorption sites on the surface and the deviation from an ideal single first-order process \citep{Doronin_2015}. The assumption of a constant prefactor over the desorption temperature range introduces a systematic uncertainty. However, this uncertainty remains small compared to the width and overall uncertainty of the adsorption energies, to which the desorption flux is exponentially sensitive \citep{Tait_2005, Minissale_2022}. Overall, two main sources of systematic uncertainty remain: the first-order approximation, which is effectively accounted for in the width of the distributions, and the second, arising from the propagation of the calibration uncertainty (about 10$\%$) into the extracted adsorption energies.

\subsection{\label{sec:result_submono}Results and discussion}

      \begin{figure}[h!]
   \centering
   \includegraphics[width=\hsize]{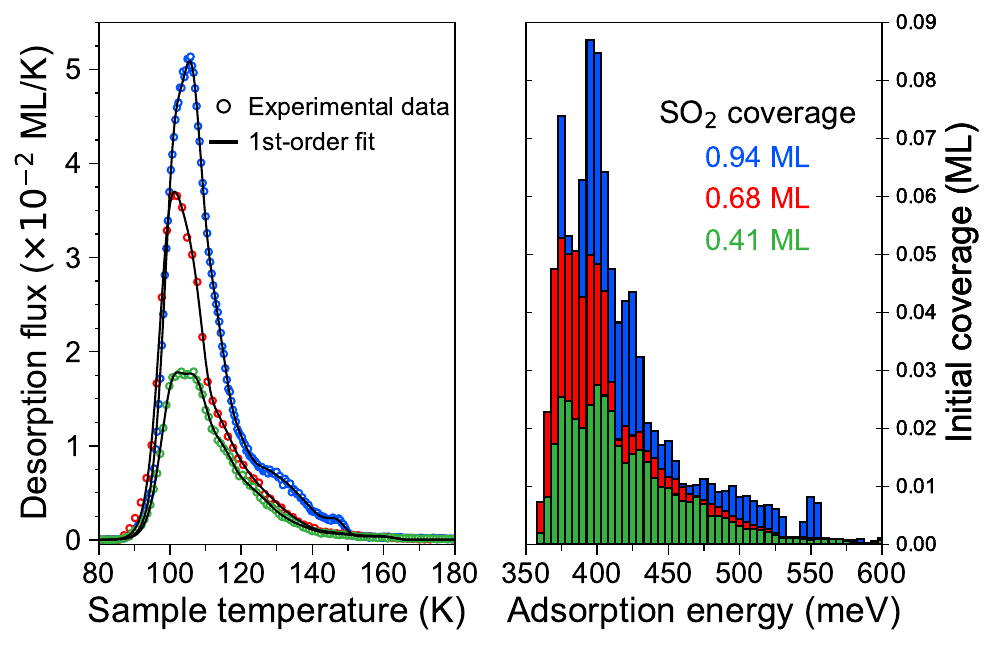}
      \caption{(Left) TPD curves of SO$_2$ deposited at 80~K on gold (circles) for different coverages (0.94~ML, blue; 0.68~ML, red; and 0.41~ML, green) in the submonolayer regime, with corresponding first-order fits (black lines). (Right) Associated adsorption energy distributions for each TPD curve.}
         \label{fig:dist_fit_so2_au}
   \end{figure}

      \begin{figure}[h!]
   \centering
   \includegraphics[width=\hsize]{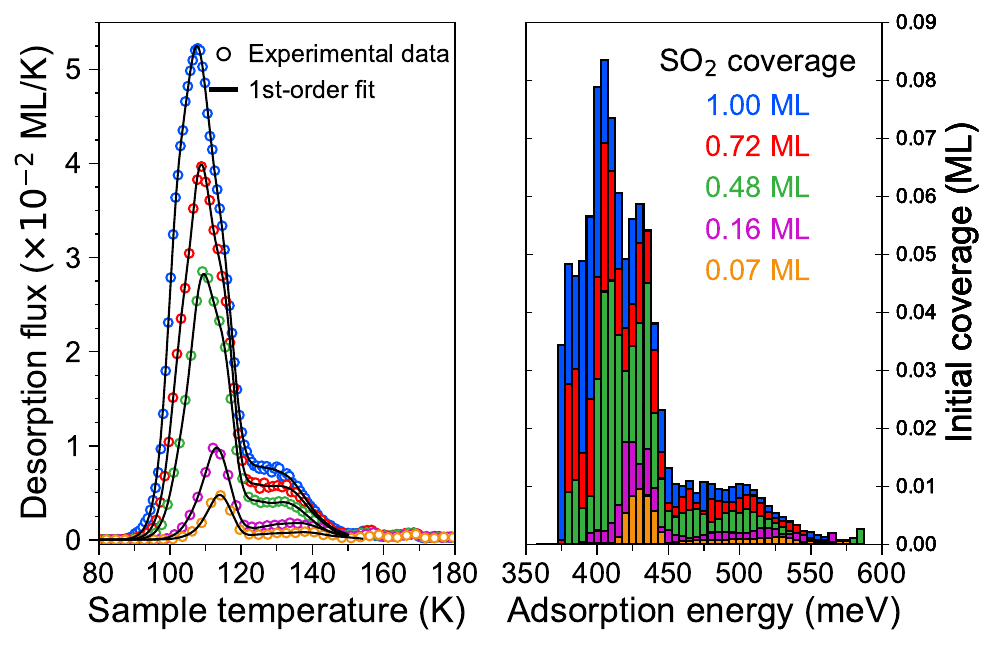}
      \caption{(Left) TPD curves of SO$_2$ deposited at 80~K on c-ASW (circles) for different coverages (1~ML, blue; 0.72~ML, red; 0.48~ML, green; 0.16~ML, magenta; and 0.07~ML, orange) in the submonolayer regime, with corresponding first-order fits (black lines). (Right) Associated adsorption energy distribution for each TPD curve.}
         \label{fig:dist_fit_so2_au_h2o}
   \end{figure}

\begin{figure}[h!]
   \centering
   \includegraphics[width=\hsize]{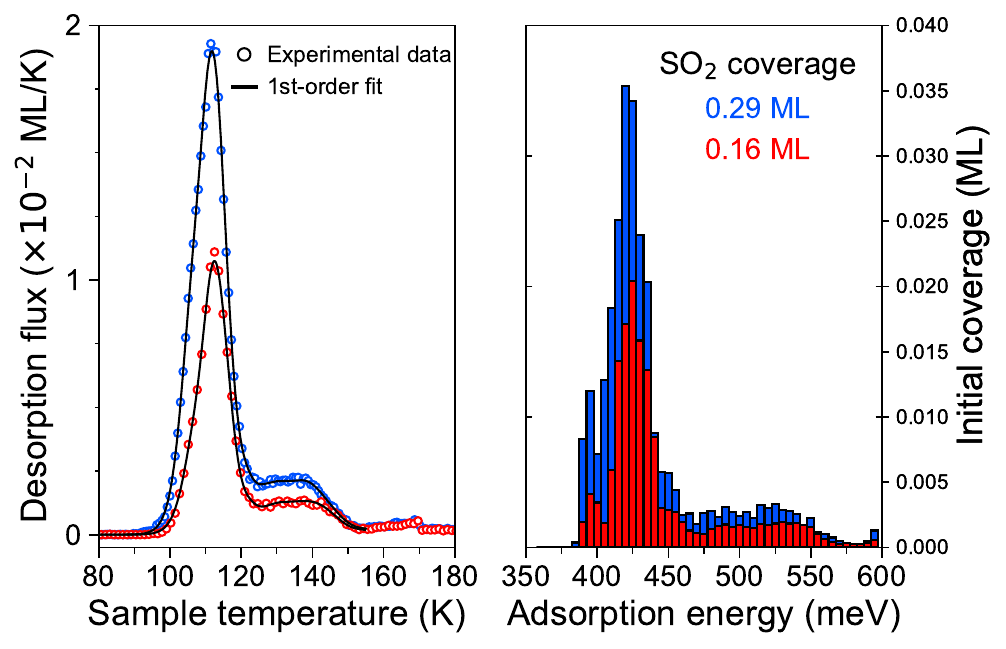}
      \caption{(Left) TPD curves of SO$_2$ deposited at 80~K on crystalline water (circles) for different coverages (0.29~ML, blue; 0.16~ML red) in the submonolayer regime, with corresponding first-order fits (black lines). (Right) Associated adsorption energy distribution for each TPD curve.}
         \label{fig:dist_fit_so2_h2o_cryst}
   \end{figure}

Figures~\ref{fig:dist_fit_so2_au}, \ref{fig:dist_fit_so2_au_h2o}, and \ref{fig:dist_fit_so2_h2o_cryst} present the results of the fits performed for the TPD curves of SO$_2$ on gold, c-ASW, and crystalline water ice, respectively.
The right panels show the corresponding adsorption energy distributions.
As shown, the fits reproduce the experimental data (represented by circles; one out of every four points is displayed for clarity) with excellent accuracy, confirming the validity of the first-order kinetic model and the use of the prefactor derived from the TST.
Indeed, modifying the prefactor by even one order of magnitude prevents the fits from converging or significantly worsens their quality.
From these distributions, we derived the most probable adsorption energy, as well as the mean and standard deviation values, summarised in Table~\ref{table:res_fit}.

The obtained energy distributions are systematically shifted towards higher adsorption energies compared to those extracted in the multilayer regime.
This confirms that zeroth-order desorption plays little to no role in this regime, as expected for submonolayer coverages.
For SO$_2$ deposited on gold (Fig.~\ref{fig:dist_fit_so2_au}), the distributions consistently peak around 400~meV before decreasing smoothly to zero.
For SO$_2$ deposited on both amorphous and crystalline water ice (Figs.~\ref{fig:dist_fit_so2_au_h2o} and \ref{fig:dist_fit_so2_h2o_cryst}), the distributions exhibit similar shapes and values at comparable coverages.
As in the gold case, a sharp maximum at approximately 400~meV is followed by rapid decay at around 450~meV, with a second distinct shoulder or bump systematically appearing around 510~meV. Moreover, in Fig.~\ref{fig:dist_fit_so2_au_h2o} (left panel), the TPD curves corresponding to the three highest initial coverages (1.00, 0.72, and 0.48~ML) clearly exhibit two components within the main desorption feature — one around 410~meV and another near 430~meV.
The first component progressively disappears as the initial coverage decreases (0.16 and 0.07~ML), indicating that it is associated with adsorption sites that become less populated at low surface coverages. We observe this double-peak structure in the desorption flux profiles of Fig.~\ref{fig:comp_SO2_cASW}. This higher-energy component provides further evidence for a specific SO$_2$–H$_2$O interaction. In all cases, both the most probable and mean SO$_2$ adsorption energies are higher when the molecule is deposited on water compared to gold.
This confirms that SO$_2$ interacts more strongly with H$_2$O, whereas the ice morphology (amorphous vs crystalline) only minimally affects the overall adsorption energy distribution and, consequently, on the adsorption process. Comparing our results with the calculations by \citep{Nguyen_2024}, we find that the mean binding energy of SO$_2$ on c-ASW is 439 $\pm$ 41 meV, which is consistent with the value of 450 meV reported in this work.

\begin{table}[h!]
\caption {Numerical results of the adsorption energy distribution.}
\label{table:res_fit}  
\centering
\begin{tabular}{cccc}
\hline\hline   
\rule{0pt}{1.em}
$\theta_0 (\mathrm{SO}_2)$ & $E_\mathrm{ads}^\mathrm{max}$ & $\langle E_\mathrm{ads} \rangle$ & $\sigma_E$\\
\hline
  \multicolumn{4}{c}{\it Substrate : Au}\\ 
\hline\rule{0pt}{1.1em}
  0.94 & 395 & 420 & 43 \\
  0.68 & 375 & 412 & 43 \\
  0.41 & 400 & 422 & 45 \\
\hline
  \multicolumn{4}{c}{\it Substrate : c-ASW} \\
\hline
  1.00 & 405 & 424 & 38 \\
  0.72 & 410 & 430 & 39 \\
  0.48 & 410 & 434 & 39 \\
  0.16 & 430 & 451 & 43 \\
  0.07 & 430 & 456 & 44 \\
\hline
\multicolumn{4}{c}{\it Substrate : crystalline H$_2$O} \\
\hline
  0.29 & 425 & 439 & 41 \\
  0.16 & 425 & 446 & 44 
\end{tabular}
\tablefoot{ All coverage are express in ML meanwhile energies are express in meV.}
\end{table}

Finally, as indicated in Table~\ref{table:res_fit}, the initial coverage and the mean adsorption energy on c-ASW appear inversely correlated.
This suggests that, as coverage decreases, SO$_2$ molecules preferentially occupy deeper or more strongly bound sites. This shift to higher adsorption energies with decreasing coverage is expected and has been observed in other systems \citep{amiaud_2006, Zubkov_2007, bertin_nitrile_2017, Minissale_2022}. Indeed, at lower coverage, the initially deposited molecules can diffuse to more strongly bound sites before desorbing during warm-up. At higher coverage, however, the less strongly bound molecules cannot diffuse to other sites as they are already occupied.
     
Interestingly, the secondary maximum observed around 500~meV in the adsorption energy distributions for SO$_2$ on amorphous and crystalline water ice coincides with the reported adsorption energy of pure H$_2$O ($\approx$ 520~meV) (\citealt{Fraser_2001}, \citealt{Speedy_1996}). This overlap suggests that this high-energy component may be related to processes occurring at temperatures close to the onset of water desorption. In such cases, it is commonly attributed to the release of molecules associated with the first layers of water ice. In this context, it is not possible -- based on TPD experiments alone -- to determine whether the extracted high adsorption energies originate from H$_2$O desorption or real SO$_2$ adsorption energies. As discussed in Sect. 3, \citep{Bang_2017} suggest that this high-energy contribution arises from SO$_2$ molecules that have thermally reacted with the water surface.

\section{Conclusion}
   
We studied adsorption and thermal desorption of SO$_2$ deposited on gold and water ice substrates. On the one hand, thermal desorption of SO$_2$ multilayers follows zero-order kinetics, allowing the extraction of both pre-exponential factors and adsorption energies. On the other hand, thermal desorption of SO$_2$ submonolayers on polycrystalline gold and water ice substrates exhibits more complex kinetics, approximated by a series of first-order kinetics, each associated with a given adsorption energy. This has enabled the extraction of SO$_2$ adsorption energy distributions on the considered substrates, using the pre-exponential factor determined by the TST \citep{Tait_2005, Minissale_2022}. The adsorption energy distributions of SO$_2$ submonolayers on both amorphous and crystalline water ice surfaces exhibit two components: a physisorbed layer with adsorption energies between 375 and 455~meV, and a more strongly bound population with adsorption energies peaking around 500~meV, indistinguishable from the supporting water desorption threshold. This second contribution is associated with SO$_2$ molecules that thermally react with the water ice surface between 100 and 120~K \citep{Bang_2017}. For compact amorphous water ice, small fractions of SO$_2$ were also shown to thermally diffuse into the ice bulk, causing volcano and co-desorption phenomena at higher temperatures (> 150~K). These contributions are, however, negligible compared to surface SO$_2$ adsorption. Otherwise, adsorption energies and behaviours on compact amorphous and on crystalline water ices show minimal differences.

Our adsorption energy values typically range from 375 and 500~meV, closely matching the cohesion energy of pure water ice. This compares well with the theoretical prediction of \cite{Nguyen_2024}, although their study highlights slightly more strongly bound sites for crystalline ice for which we find no evidence. However, this is significantly higher than the value of 260~meV proposed by \cite{penteado_sensitivity_2017}, which is based on a rough estimation of the mean desorption temperatures from \cite{Collings_2004}. This also suggests that SO$_2$ sublimation from the grain surface occurs simultaneously with that of water and, thus, the icy mantle itself. This supports the fact that gaseous SO$_2$ is a good tracer for hot cores, such as recently proposed by mid-infrared and millimetre observations \citep{van_gelder_joys_2024}.
More generally, our study provides adsorption energy distributions -- including mean values, most probable values, and distribution sizes -- that can be directly incorporated into astrochemical models accounting for both gaseous and solid phases, as well as gas-grain exchanges. These values are central for modelling solid-to-gas abundance ratios with temperature and for estimating the incorporation, diffusion, and even chemistry of SO$_2$ on water-rich ices.

\begin{acknowledgements}
      This work was supported by the Programme National Physique et Chimie du Milieu Interstellaire (PCMI) of CNRS/INSU with INP/INP cofunded by CEA and CNES. A.H. acknowledges PhD funding by the DIM-ORIGINES (IDF-DIM-ORIGINES-2022-1-02) program of the Region Iles-de-France. 
\end{acknowledgements}

\bibliographystyle{aa}
\bibliography{biblio.bib}
\end{document}